\documentclass[reprint,aps,prb]{revtex4-1}
\usepackage{color,graphicx,calc,url}
\usepackage{dcolumn}
\usepackage{bm}
\usepackage{amssymb}
\usepackage{amsmath}
\usepackage{wasysym}
\usepackage{mathrsfs} 

\usepackage{graphicx}
\usepackage[version=3]{mhchem}
\usepackage{siunitx}

\begin{document}
\title{Co-sputtered PtMnSb thin films and PtMnSb/Pt bilayers for spin-orbit torque investigations}
\author{Jan Krieft$^1$, Johannes Mendil$^2$, Myriam H. Aguirre$^{3,4}$, Can O. Avci$^5$, Christoph Klewe$^6$, Karsten Rott$^1$, Jan-Michael Schmalhorst$^1$, G\"unter~Reiss$^1$, Pietro Gambardella$^2$ and Timo~Kuschel$^{1,7}$\email{Electronic mail: jkrieft@physik.uni-bielefeld.de}}
\affiliation{$^1$Center for Spinelectronic Materials and Devices, Department of Physics, Bielefeld University, 33615 Bielefeld, Germany\\
$^2$ Department of Materials, ETH Z\"urich, H\"onggerbergring 64, CH-8093 Z\"urich, Switzerland\\
$^3$ Departamento de Fisica de la Materia Condensada, Universidad de Zaragoza, E-50009 Zaragoza, Spain\\
$^4$ Instituto de Nanociencia de Arag{\'o}n (INA) \& Laboratorio de Microscop{\'i}as Avanzadas (LMA), Universidad de Zaragoza, E-50018 Zaragoza, Spain\\
$^5$ Department of Materials Science and Engineering, Massachusetts Institute of Technology, Cambridge, Massachusetts 02139, USA\\
$^6$ Advanced Light Source, Lawrence Berkeley National Laboratory, Berkeley, California 94720, USA\\
$^7$ Physics of Nanodevices, Zernike Institute for Advanced Materials, University of Groningen, 9747 AG Groningen, The Netherlands}

\date{\today}


\begin{abstract}
The manipulation of the magnetization by spin-orbit torques (SOTs) has recently been extensively studied due to its potential for efficiently writing information in magnetic memories. Particular attention is paid to non-centrosymmetric systems with space inversion asymmetry, where SOTs emerge even in single-layer materials. The half-metallic half-Heusler \ce{PtMnSb} is an interesting candidate for studies of this intrinsic SOT. Here, we report on the growth and epitaxial properties of \ce{PtMnSb} thin films and \ce{PtMnSb}/\ce{Pt} bilayers deposited on \ce{MgO}(001) substrates by dc magnetron co-sputtering at high temperature in ultra-high vacuum. The film properties were investigated by x-ray diffraction, x-ray reflectivity, atomic force microscopy, and electron microscopy. Thin \ce{PtMnSb} films present a monocrystalline C1b phase with (001) orientation, coexisting at increasing thickness with a polycrystalline phase with (111) texture. Films thinner than about \SI{5}{nm} grow in islands, whereas thicker films grow layer-by-layer, forming a perfect \ce{MgO}/\ce{PtMnSb} interface. The thin \ce{PtMnSb}/\ce{Pt} bilayers also show island growth and a defective transition zone, while thicker films grow layer-by-layer and \ce{Pt} grows epitaxially on the half-Heusler compound without significant interdiffusion.

\end{abstract}

\maketitle


\begin{figure*}
\centering
\includegraphics[width=\linewidth]{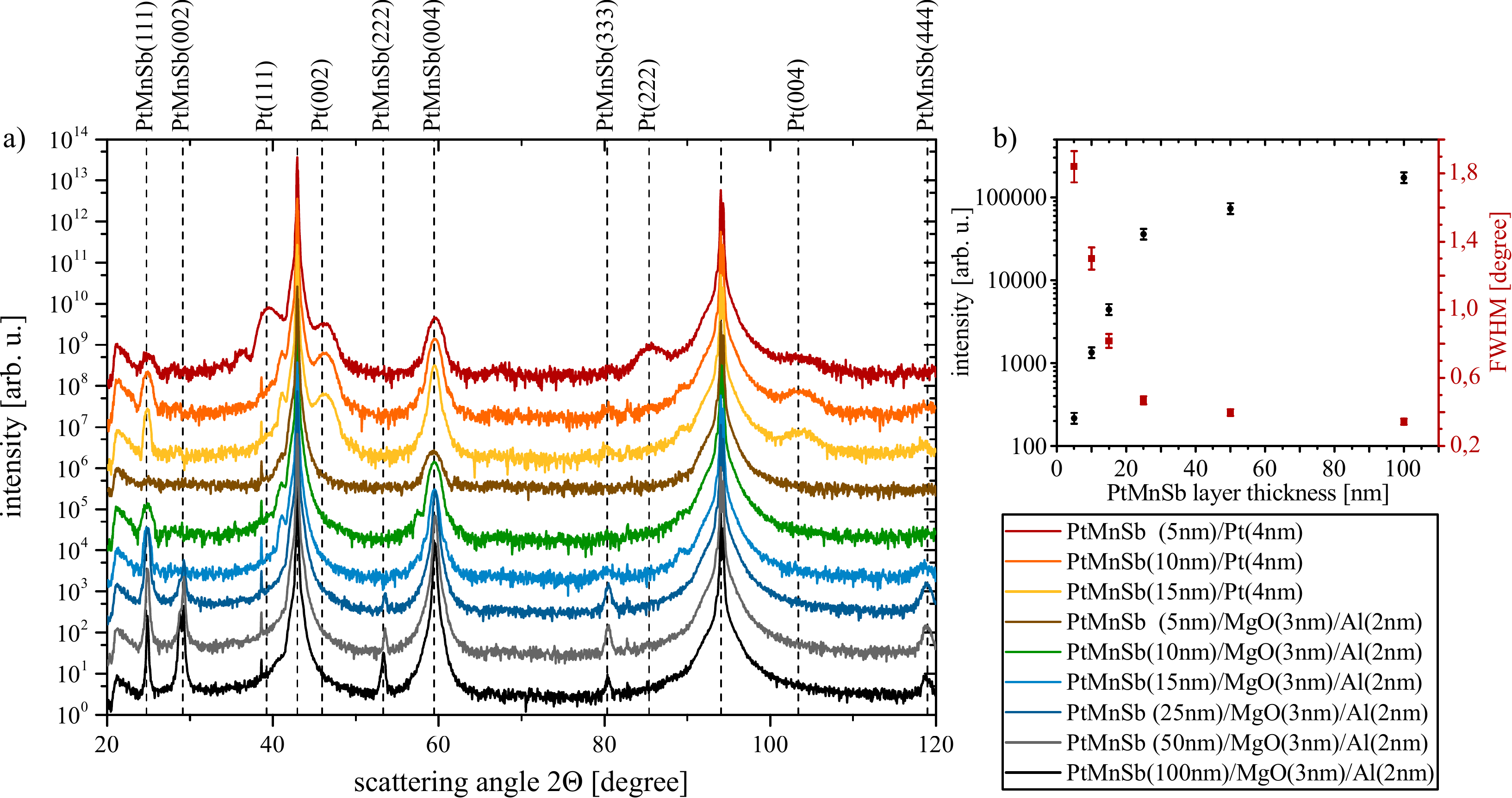}
\caption{a) XRD measurements including \ce{PtMnSb}(111) and \ce{PtMnSb}(004) Bragg reflections indicating two domain orientations. b) Bragg peak intensity and FWHM of the \ce{PtMnSb}(004) reflection plotted against the \ce{PtMnSb} layer thickness of the \ce{MgO}/\ce{Al} capped samples.} 
\label{fig:XRD1}
\end{figure*} 

Spin-orbit torques (SOTs) in ferromagnet/heavy metal bilayers can be utilized to electrically manipulate and switch the magnetization in the magnetic layer.\cite{miron2010current, miron2011perpendicular, liu2012spin, neumann2016temperature, avci2016current} In these systems, the spin Hall effect \cite{dyakonov1971current, sinova2015spin} in the heavy metal layer and electron scattering at the interface\cite{PhysRevB.94.104420} induce a spin current from the heavy metal to the adjacent magnetic layer, which gives rise to SOT. This has been demonstrated for ferromagnetic semiconductors\cite{chernyshov2009evidence} and, more recently, for the room-temperature ferromagnet half-Heusler compound \ce{NiMnSb}\cite{ciccarelli2016room} and the antiferromagnet \ce{CuMnAs}.\cite{wadley2016electrical} The family of half-metal Heusler alloys with C1b structure is particularly interesting for SOT studies because of their robust magnetic properties and lack of inversion symmetry.

In this work, we prepare and investigate co-sputter-deposited thin films of the room-temperature ferromagnet \ce{PtMnSb}, which belongs to the same Heusler class of materials.\cite{watanabe1970new, de1983new}
\ce{PtMnSb} thin films were first investigated in the 1980s as magnetooptical recording media,\cite{van1983ptmnsb} due to their very large magnetooptical Kerr effect.\cite{akasaka1989process} Half-metallic alloy films have been prepared by rf sputtering from a sintered \ce{MnSb} target with \ce{Pt} sheets or from a mosaic target 
on a fused quartz or glas substrate followed by annealing.\cite{ohyama1987magneto, attaran1989magnetic}
To investigate the possibility of using the ferromagnet \ce{PtMnSb} in giant magnetoresistance applications, epitaxial \ce{PtMnSb}(111) films on \ce{Al2O3}(0001) have been produced with dc magnetron co-sputtering at \SI{500}{\celsius}.\cite{kautzky1997investigation} Epitaxial \ce{PtMnSb}(001) films were also obtained by dc co-sputtering on an \ce{MgO}(001) substrate, using an oriented \ce{W} seed layer. This growth is accompanied by a modest in-plane strain insufficient to induce perpendicular magnetization through inverse magnetostriction.\cite{matsui1993effect} Moreover, ordered thin films of \ce{PtMnSb} can be produced by solid-state phase formation via thermal annealing of compositionally modulated thin multilayered \ce{Pt}/\ce{Mn}/\ce{Sb} films. In this process, films consisting of the two phases \ce{PtMnSb} and \ce{Mn2Sb} have been obtained. Structurally related half-Heusler compounds such as \ce{NiMnSb} can equally be grown by pulsed-laser deposition using low substrate temperature. \cite{giapintzakis2002low} In addition, epitaxial \ce{NiMnSb}(001) thin films can be fabricated on (\ce{In},\ce{Ga})\ce{As}(001) by molecular-beam epitaxy (MBE).\cite{van2000epitaxial, bach2003molecular, gerhard2014control} 
\quad\newline

We prepared epitaxial \ce{PtMnSb} thin films with varying film thickness on \ce{MgO}(001) substrates at high temperatures without utilizing a seed layer  and fabricated additional \ce{PtMnSb}/\ce{Pt} bilayers in order to perform electrical measurements and ultimately study spin transport properties and SOTs.\cite{garello2013symmetry, avci2015unidirectional} Detailed x-ray technique analysis in combination with high resolution transmission electron microscopy was performed to characterize the film growth and quality.


The \ce{PtMnSb} films are prepared by dc magnetron co-sputtering from three 3-inch sources in an ultra-high vacuum chamber with a base pressure of \SI{E-9}{\milli\bar}. In each preparation cycle, two nominally identical \ce{PtMnSb}(x\,nm) layers were grown on two polished \ce{MgO}(001) substrates with a lattice mismatch of about \SI{4}{\percent}, one of them with Pt(x\,nm) on top. Sputter powers used for the elemental targets are \SI{25}{\watt} for \ce{Pt}, \SI{38}{\watt} for \ce{Mn} and \SI{41}{\watt} for \ce{Sb} with \SI{10}{sccm} \ce{Ar} gas flow at \SI{1.9E-3}{\milli\bar} to obtain the correct stoichiometry with a deposition rate of \SI{0.11}{nm/sec} for the total thickness. The sources are confocally arranged and tilted towards the substrate at an angle of \SI{35}{\degree} with respect to surface normal. The substrates are rotated during deposition at \SI{10}{rpm} to avoid inhomogeneous covering. Deposition temperature is between \SI{400}{\celsius} and \SI{475}{\celsius} to achieve optimal growth followed by a cool down to room temperature in about \SI{2}{\hour}. The structural analysis reveals that temperatures higher than \SI{400}{\celsius} are incompatible with monocrystalline growth and not suitable for further investigation of the magnetic and transport properties due to increased granular character of the films.

Subsequently, the \ce{Pt} layer is deposited at room temperature at \SI{100}{\watt} with \SI{10}{sccm} \ce{Ar} gas flow at \SI{1.9E-3}{\milli\bar} and substrate rotation, with a \ce{Pt} deposition rate of \SI{0.09}{nm/sec}. For the twin sample fabrication with and without \ce{Pt} top layer, the \ce{PtMnSb} film on one \ce{MgO} substrate is covered by a mask prior to the \ce{Pt} deposition. Afterwards and prior to atmosphere exposure, the single \ce{PtMnSb} layer is capped at room temperature by \ce{MgO}(\SI{3}{nm}) to avoid ex-situ oxidation and by \SI{2}{nm} of \ce{Al} forming a passivating oxide to protect the hygroscopic \ce{MgO} layer.

\begin{figure*}
\centering
\includegraphics[width=\linewidth]{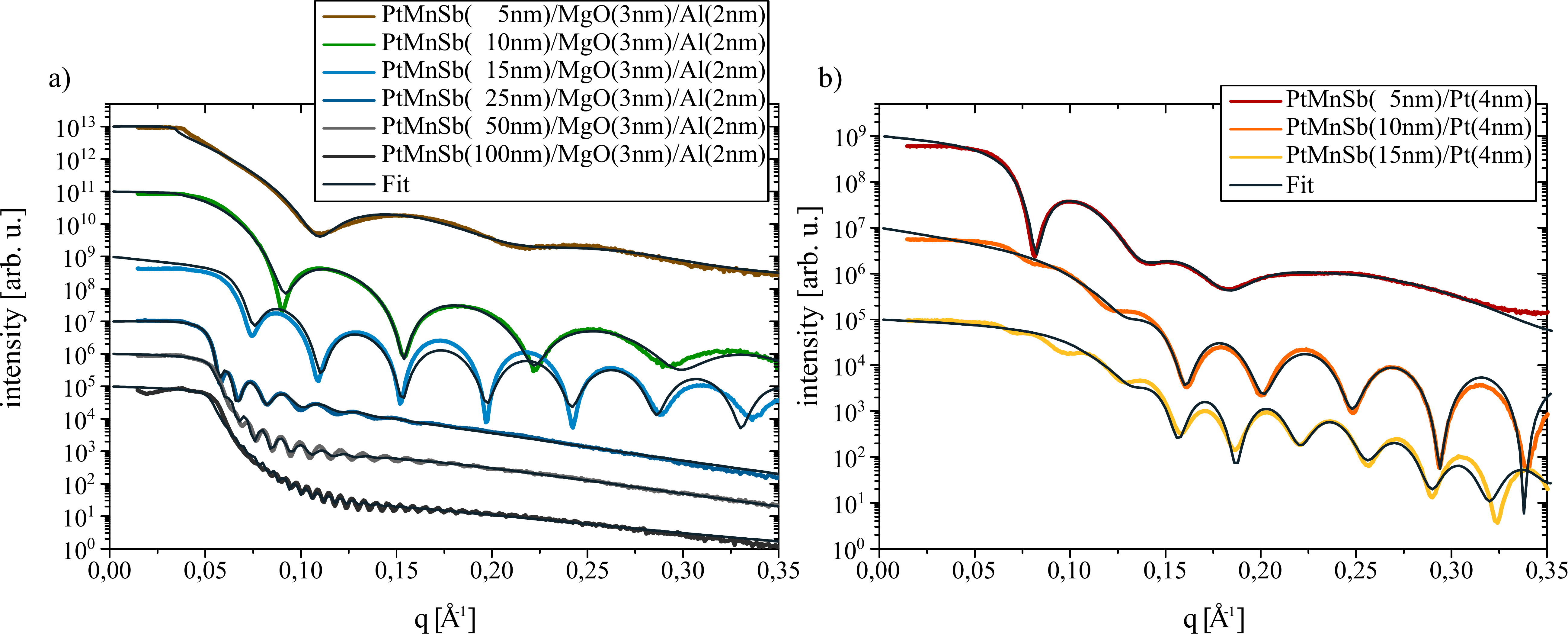}
\caption{XRR measurements of a) single \ce{PtMnSb}(x\,nm) layers with capping and b) \ce{PtMnSb}(x\,nm)/\ce{Pt} bilayers. XRR measurements show Kiessig oscillations which can be described by a typical Parratt layer-by-layer model using additional transition layers for the bilayer system.}
 
\label{fig:XRR1}
\end{figure*} 


\begin{figure}[b]
\centering
\includegraphics[width=1.0\linewidth]{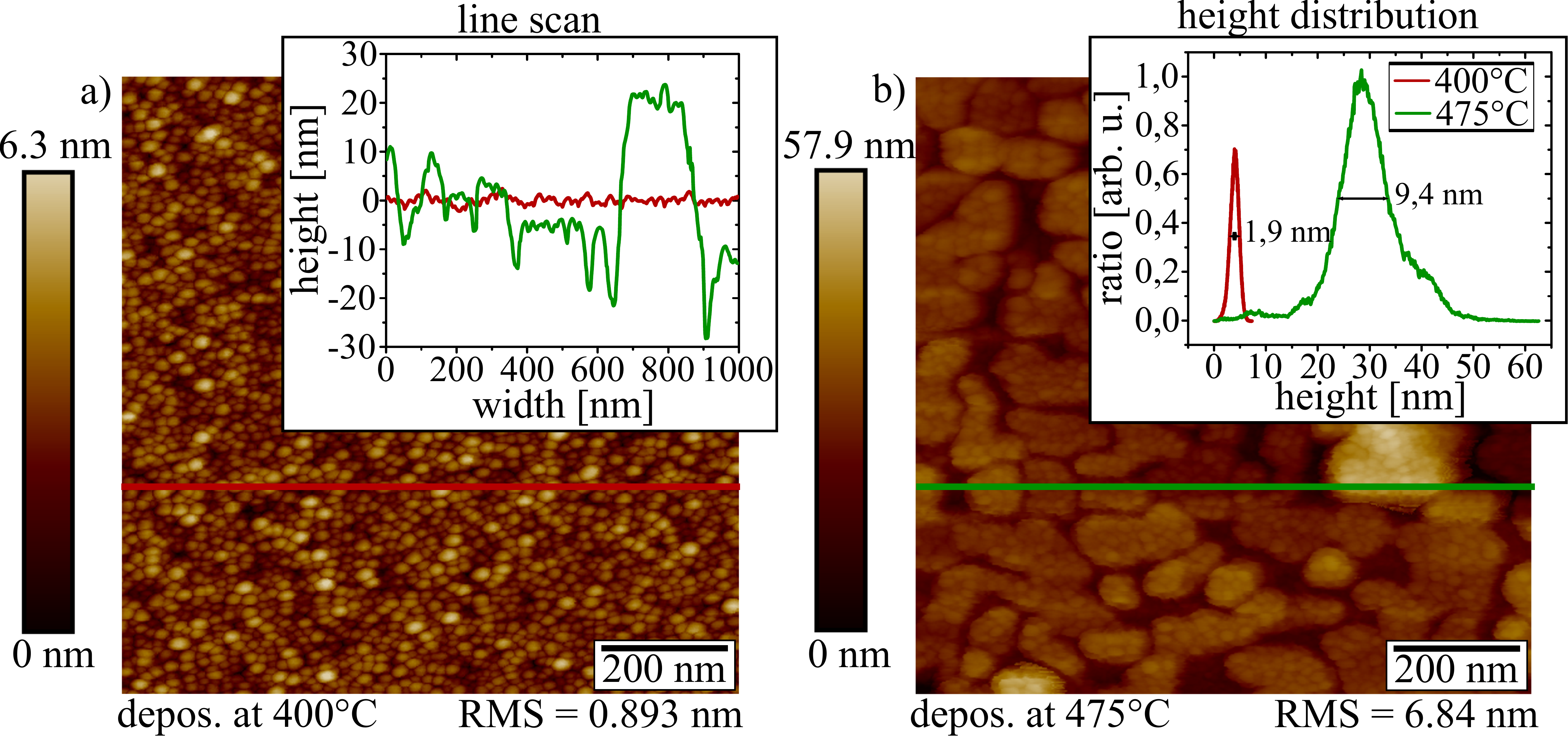}

\caption{AFM images of \SI{10}{nm} \ce{PtMnSb} films deposited at a) \SI{400}{\celsius} and b) \SI{475}{\celsius}. Higher deposition temperature indicates larger grains and more distinct island growth. The corresponding line scans and height distributions (insets) emphasize the difference in roughness and \ce{PtMnSb} film growth.}
\label{fig:AFM1}
\end{figure} 

After deposition, the composition of the films is determined by spectral energy-dispersive x-ray spectroscopy (EDX). X-ray diffraction (XRD) measurements are performed with a Philips X'Pert Pro diffractometer with a \ce{Cu} $\text{K}_{\alpha}$ source in a Bragg Brentano configuration to analyze the samples crystallographic properties. The film thickness and roughness are determined via x-ray reflectivity (XRR) in the same setup. A Bruker AFM Multimode is used for atomic force microscopy (AFM) of the surface topography utilizing Bruker FMV-A probes in tapping mode. The thin films crystallinity, structure and composition is analyzed by a scanning transmission electron microscope (STEM) with a high angular annular dark field (HAADF) detector in a FEI Titan G2 60-300 equipment with an aberration corrected probe. HAADF imaging is used to map the thin layer via STEM microscopy which is highly sensitive to element variations. The preparation of uniformly thin TEM lamellae is done by focused ion beam milling in a Dual Beam Helios 650 Nanolab. We visualize the atomic structure to check growth and layer roughness. The EDX analysis is performed by in-situ-STEM with an Oxford equipment. 


\begin{figure*}
\centering
\includegraphics[width=\linewidth]{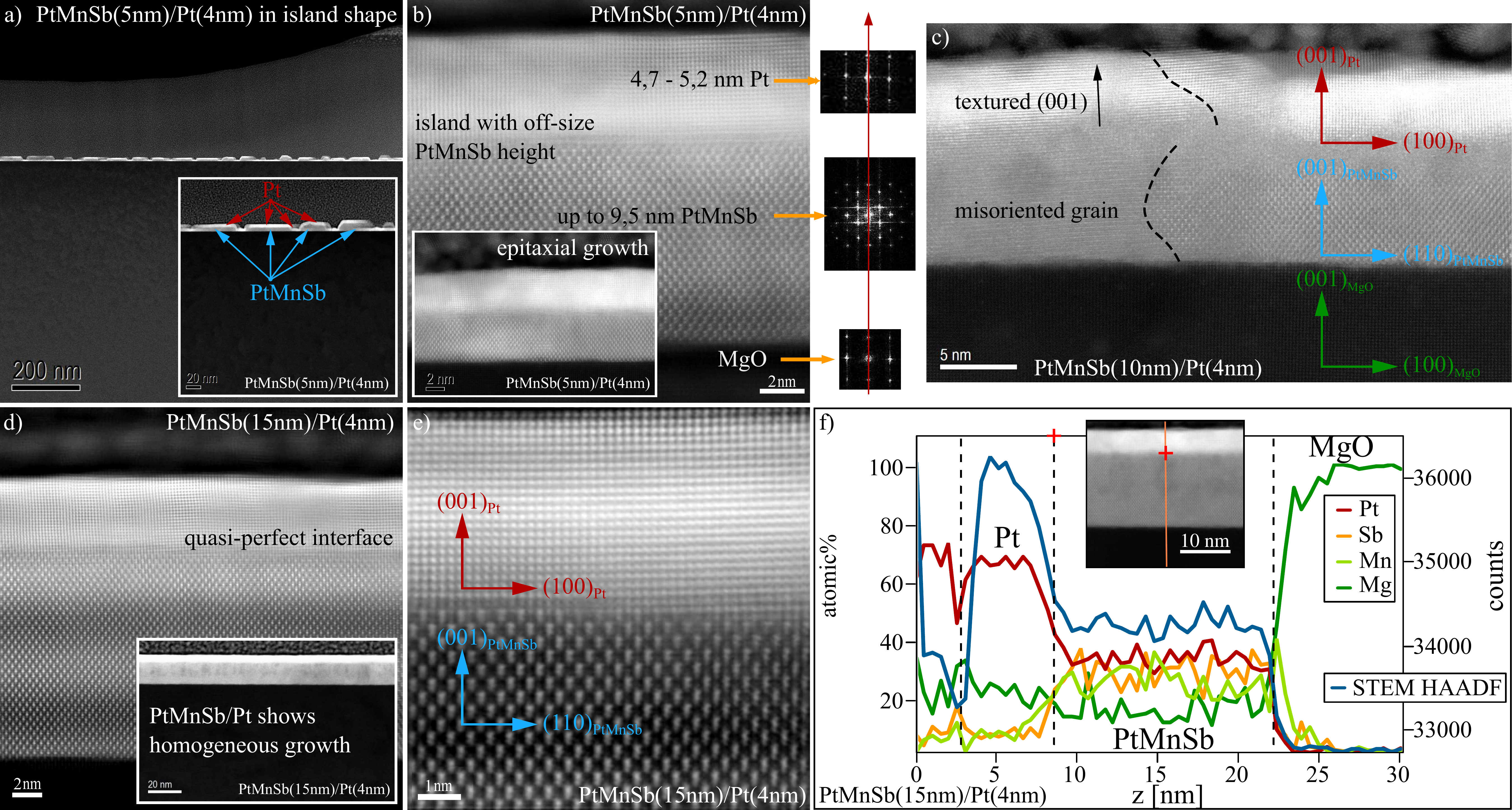}
\caption{STEM-HAADF images reveal local crystalline growth of \ce{PtMnSb} and \ce{Pt} on the \ce{MgO} substrate. a) Films with small \ce{PtMnSb} thicknesses of \SI{5}{nm} show elongated islands. Inset shows a zoom-in of this bilayer. b) STEM-HAADF image illustrates an anomaly in the \ce{PtMnSb}(5nm)/\ce{Pt}(\SI{4}{nm}) bilayer where an island grew up to \SI{9.5}{nm}. Inset shows the standard \SI{5}{nm} thin film sample epitaxial growth. Red arrow across the calculated diffraction patterns indicates a coherent axis in direction of growth. c) Thicker films of \SI{10}{nm} show a small transition zone and misoriented grains. d) Films of \SI{15}{nm} thickness have layer-by-layer growth, as seen in the inset showing homogenous growth. e) STEM-HAADF image with Wienner filtering for noise removing of the interface and the EDX linescan profile f) reveal quasi-perfect interface
matching without an interdiffusion region.}
\label{fig:TEM1}
\end{figure*} 

Figure \ref{fig:XRD1}a) shows the results of the XRD analysis. The \ce{PtMnSb}(004) and (111) Bragg peaks are clearly visible for all samples, while the (002), (222) and (333) only appear when the layer thickness exceeds about \SI{15}{nm}. Therefore, all films show at least a polycrystalline but textured structure with major epitaxial growth in the [001] direction, as can be concluded from the peak intensity ratios and complementary calculations of the structure factors. Off-specular XRD experiments using an Eulerian cradle confirm the crystalline growth of the \ce{PtMnSb} with (001) orientation, since corresponding sharp Bragg reflections can be
detected. Off-specular Bragg reflections from the (111) orientation could not be found. Therefore, the specular
\ce{PtMnSb}(111) reflections belong to polycrystalline material with (111) texture, while the dominant \ce{PtMnSb}(004) reflection is due to a monocrystalline \ce{PtMnSb} phase. Figure \ref{fig:XRD1}b) shows the (004) Bragg peak intensity and full width at half maximum (FWHM) plotted against the \ce{PtMnSb} film thickness. The flattening of the intensity increase between \SI{15}{nm} and \SI{25}{nm} points to an increasingly polycrystalline growth accompanied by a reduction of texture. Due to direct growth on the \ce{MgO} substrate, for thin \ce{PtMnSb} layers epitaxial growth of \ce{Pt} in (111) and (001) orientation can be observed.  With increasing thickness \ce{Pt} is growing quasi-epitaxially on the ordered \ce{PtMnSb} in (001) orientation, hence only the \ce{Pt}(002) and (004) peaks are visible (cf. Fig.\,\ref{fig:TEM1}e)).

The pronounced oscillations in the XRR scans shown in Fig.\,\ref{fig:XRR1} verify relatively smooth surfaces for all layers exceeding \SI{5}{nm} with roughnesses in the range of \SI{0.7}{nm} to \SI{1.2}{nm}. This nicely fits the roughness value obtained by additional AFM measurements shown in Fig.\,\ref{fig:AFM1}. The $\SI{1}{\micro\meter} \times \SI{1}{\micro\meter}$ scan reveals a locally smooth and homogeneous growth with granular character for thin films prepared at \SI{400}{\celsius} (cf. Fig.\,\ref{fig:AFM1}a)). We find larger grains and a more distinct island growth for an increasing deposition temperature of \SI{475}{\celsius} as exemplarily shown in Fig.\,\ref{fig:AFM1}b). The depth profile of the dispersion $\delta$ is simulated for the XRR fits by a layer-by-layer model which depends on the Heusler compound thickness and growth. 
Consequently, the density distribution is significantly affected by the island growth of \ce{PtMnSb} (cf. TEM results in Fig.\,\ref{fig:TEM1}a)) and the thickness dependent-transition zone between \ce{PtMnSb} and \ce{Pt}.

Figure 4 shows representative STEM images of \ce{PtMnSb}/\ce{Pt} bilayers for different thicknesses of \ce{PtMnSb}. We find that the thinnest layer, \ce{PtMnSb}(\SI{5}{nm})/\ce{Pt} (cf. Fig.\,\ref{fig:TEM1}a) and b)) grows epitaxially in a Volmer-Weber mode forming a sharp interface with MgO. Dark field images show the non-wetting elemental micro and granular structure of the \ce{PtMnSb}. Structural coherence and epitaxy is found in large regions of \SI{20}{}-\SI{60}{nm}. In bilayers the \ce{Pt} overlayer grows uniformly over the substrate, covering both \ce{PtMnSb} islands and the bare \ce{MgO} substrate in a uniform way. Therefore, in regions without \ce{PtMnSb}, the \ce{Pt} is growing directly on the \ce{MgO}. The interface of \ce{PtMnSb}(\SI{5}{nm})/\ce{Pt} is partially defective, as seen in figure \ref{fig:TEM1}b), showing a STEM-HAADF image of an island of the nominal \SI{5}{nm} film with a height of up to \SI{9.5}{nm}.

The intermediate thickness layer, \ce{PtMnSb}(\SI{10}{nm})/\ce{Pt} (cf. Fig.\,\ref{fig:TEM1}c)), shows a much smaller roughness and relatively homogeneous growth. A small transition zone manifests at the interface, where the \ce{Pt} is forced into a disordered cubic structure. STEM-HAADF images indicate partially epitaxial films with a transition zone adjusting to both crystal structures. Hence, these images reveal a defective quasi-amorphous interface between \ce{Pt} and \ce{PtMnSb} in half-Heusler structure. Areas with excess \ce{Pt} tend to grow adopting the fcc structure instead of the half-Heulser structure. In these intermediate layers we find misoriented grains, exemplary illustrated on the right side of Fig.\,\ref{fig:TEM1}c), where the lattice structure of \ce{PtMnSb} cannot be resolved.

Finally, the thickest \ce{PtMnSb}(\SI{15}{nm})/\ce{Pt} layer (cf. Fig.\,\ref{fig:TEM1}d) and e)) exhibits layer-by-layer growth. The interface only shows some steps and dislocations for fitting the two lattices with an almost perfect match (cf. Fig.\,\ref{fig:TEM1}e)). Consequently, we find that thin \ce{PtMnSb} layers exhibit island growth and a defective \ce{PtMnSb}/\ce{Pt} transition zone, while thicker films show homogeneous epitaxial growth and high quality interfaces. Figure \,\ref{fig:TEM1}f) shows the transversal EDX linescans with drift correction, confirming the composition of the \ce{PtMnSb}/\ce{Pt} bilayer films.

In conclusion, we have prepared highly textured \ce{PtMnSb} thin films and \ce{PtMnSb}/\ce{Pt} bilayers on \ce{MgO}(001) substrates by co-sputter deposition. \ce{PtMnSb} films grow epitaxially with a thickness-dependent degree of coherence and texture. The crystal structure corresponds to that of a C1b Heusler alloy. In bilayer systems, \ce{Pt} grows as a uniform polycrystalline fcc layer, in which the interface quality is significantly improving with the \ce{PtMnSb} thickness. While thin \ce{PtMnSb} films of \SI{5}{nm} grow as separate islands, thicker films such as \SI{15}{nm} and above show layer-by-layer growth. The \ce{Pt} covers the \ce{PtMnSb} and the \ce{MgO} substrate between the islands on the \ce{PtMnSb}(\SI{5}{nm}) thin films, whereas it grows epitaxially and flat on \ce{PtMnSb}(\SI{15}{nm}) films. The interface quality highly increases with the thickness of \ce{PtMnSb}, up to quasi-perfect matching. Locally, the \ce{Pt} grows crystalline on the \ce{PtMnSb} as seen in the STEM images, but on average the XRD analysis reveals the polycrystalline textured character for the whole \ce{PtMnSb}/\ce{Pt} film. While \ce{PtMnSb} has a monocrystalline phase with (001) orientation deduced from off-specular XRD measurements, a second polycrystalline phase with (111) texture is present. These results show that heterostructures based on the \ce{PtMnSb} Heusler alloy are suitable to realize spintronic devices and study SOT in noncentrosymmetric systems.

\bibliography{bibliography}

\end{document}